# Flow Fusion – Exploiting Measurement Redundancy for Smarter Allocation

Christine Foss Sjulstad, Solution Seeker AS
Danielle Monteiro, Solution Seeker AS
Bjarne Grimstad, Solution Seeker AS

## 1    INTRODUCTION

To efficiently operate a petroleum production system on a daily basis it is crucial to gather information about the actual flow rates.  This information can act as the foundation for short-term production optimization and reservoir management, in addition to aiding a production engineer in their routines within monitoring and flow assurance.  Moreover, flow rate estimates or measurements are often applied in allocation, where the objective is to accurately distribute the total hydrocarbon production to the contributing streams, such as the individual wells [1]-[2].

Due to operators sharing production facilities, pipelines and export terminals, the production systems grow more and more complex in line with the construction of new tie-ins and wells, and production allocation is becoming increasingly challenging [3].  To mitigate the potential misallocation of production in such complex systems, it has previously been essential to equip these with additional instrumentation at measurement points that have been lacking high-quality instrumentation.  In fact, some points might not even have instrumentation.  However, to implement flow metering equipment at all valuable positions in the production system is, in most cases, neither physically nor economically feasible [4]-[5].  Fortunately, the oil and gas sector has also experienced an upsurge of technological developments within data-driven and machine learning-based virtual flow metering through recent years, albeit the adoption rate is still quite low [6].  In fact, most virtual flow meters in use today are mechanistic, such as Prosper, LedaFlow, Olga and FlowManager.  An extensive review of these can be found in [7].  Virtual instrumentation supplements the physical flow meters and increases measurement redundancy.

Flow rate measurements are, like any measurement, uncertain and of varying quality [8]-[9]. According to [4], this can be considered a fundamental issue for allocation.  When the flow rate measurements are faulty, they may, in sum, deviate from the corresponding topside or export rate, giving rise to a mass imbalance in the system.  Ensuring that all flow meters throughout the production system produce high accuracy measurements is a costly affair, and perhaps an impossible one at times [5].  Hence, in practice, erroneous measurements result in production misallocation, thereby causing a potentially unjust revenue split between stakeholders [2].  Furthermore, other negatively affected fields of practice include reservoir simulation, drawdown strategy, and infill campaigns.  From this comes the increasing demand for more sophisticated, digital solutions that can mitigate measurement errors and optimize the production allocation for more complex production systems [10].

Conventional proportional allocation methods in the industry, that are referred to in more detail in [2], are limited in terms of their ability to correct systematic errors, also known as gross errors, that arise from imprecise modelling, drifting measurement errors or equipment failures and often lead to bias in the

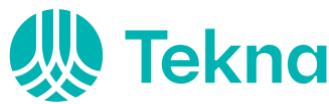 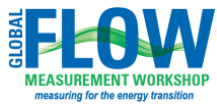 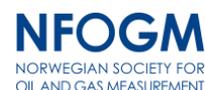

Presented at Global Flow Measurement Workshop | Tønsberg, Norway | 2023

measurements [8]-[9]. As the error is naively distributed to all sensors in the system instead of attributed to the faulty ones, the proportional allocation methods frequently give biased results. *Data validation and reconciliation* (DVR) is a more sophisticated alternative that can handle this type of error more appropriately. The method exploits measurement redundancy to minimize random measurement errors and considers the uncertainty of the overall system to detect systematic errors. With *gross error detection* (GED) algorithms in place, measurements that exhibit gross error behavior can be filtered out from the problem and the optimization can be rerun to produce more unbiased results, albeit requiring data redundancy to do so. Although the primary objective of the gross error detections is to derive a sound and justifiable allocation of the production, it can be argued that they are just as valuable in and of themselves, facilitating efficient routines within condition-based monitoring, well testing, and so on.

In this paper, we introduce a field-proven framework for reconciliation and allocation based on flow fusion - sensor fusion for flow rates. The framework is built on top of the DVR methodology and consists of four modules: data processing, uncertainty estimation, reconciliation, and GED. In this paper, we delve into the latter module and show how the solution, when data redundancy is present, offers detections of measurements that are performing out-of-spec to potentially remove them altogether from the reconciliation problem. In the following, we first present the theory behind DVR and GED, before the allocation problem within petroleum production is described alongside the different types of measurements that can be applied in an allocation framework. Thereafter, we demonstrate some examples of GED from a real-life application of the technological framework and discuss their significance in context of the measurements' nature. Lastly, some final remarks conclude the paper.

## 2      DATA VALIDATION AND RECONCILIATION

### 2.1    The methodology

DVR is an optimization-based allocation method that exploits process data redundancy to minimize measurement errors across the production system [10]. The methodology can be applied for any industrial process, such as that of chemical plants, power plants and petroleum production systems that have several measurement locations in their system. These systems are often hierarchical and can be described in terms of their M nodes and N measurements, exemplified by the simplified structure shown in Fig. 1.

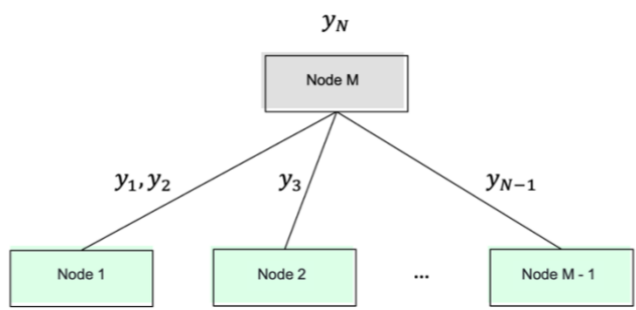

Fig. 1 - Schematic representation of a single-tier process system

DVR is a highly versatile data fusion framework that supports asymmetrical measurement setups, like the one shown in Fig. 1, and multi-tier systems. It considers the uncertainties of the individual measurements to appropriately redistribute an imbalanced material flow. The main assumption of DVR is that there are no systematic errors present in the set of measurements - an assumption that is frequently invalid in most process industries. Even after measurements are calibrated to reference values there often remains some biases. Then GED can be advantageously applied to identify the biased measurements and remove them from the reconciliation to mitigate misallocation. The steady-state DVR problem can be expressed as follows:

$$y = Cy^* + \epsilon, \epsilon \sim N(0, \Sigma) \quad (1)$$

In Equation (1), $y^* = (y_1^*, \ldots, y_M^*)$ is a vector of the true values at the $M$ nodes, $y = (y_1, \ldots, y_N)$ is a vector of the $N$ measurements, and $\epsilon = (\epsilon_1, \ldots, \epsilon_N)$ is a vector of random measurement noise. $C \in \{0,1\}^{N \times M}$ is a binary measurement matrix. Typically, the measurements are assumed to be independent of each other such that the covariance of the measurement noise can be defined as $\Sigma = diag(\sigma_1^2, \ldots, \sigma_N^2)$. The measurements follow a normal distribution $y \sim N(Cy^*, \Sigma)$, and the DVR problem can be derived by maximizing the likelihood function of this normal distribution. It takes the form of a constrained least-squares optimization problem [8]:

$$\hat{y} = \arg\min_{y^*} \{(y - Cy^*)^T \Sigma^{-1}(y - Cy^*) : Ay^* = 0\} \quad (2)$$

$A \in \{-1, 0, 1\}^{Q \times M}$ is the matrix that incorporates $Q$ system constraints. In the system depicted in Fig. 1, one typical constraint would be: $y_M^* = \sum_{i=1}^{M-1} y_i^*$, which ensures that a mass balance is maintained. The steady-state DVR problem in Equation (2) can be solved analytically, where the solution is given by:

$$\hat{y} = Ry \quad (3)$$

where $R = V(I - A^T(AVA^T)^{-1}AV)C^T\Sigma^{-1}$ and $V = (C^T\Sigma^{-1}C)^{-1}$.

### 2.2 Gross error detection

As mentioned in the previous sections, although DVR assumes that the measurements are without systematic error, all measurements in industrial processes are subject to errors of both random and systematic character [8]-[9]. The latter can arise from modelling errors, drifting measurement errors, or equipment failures. Regardless of their source, systematic errors undermine the statistical foundation of the DVR methodology and cause bias in the sensor signals, thereby introducing bias into the reconciled values themselves [11]. According to [12], this motivates for the detection and elimination of gross errors as a means to achieve a valid reconciliation - a process that could potentially follow the example workflow shown in Fig. 2.

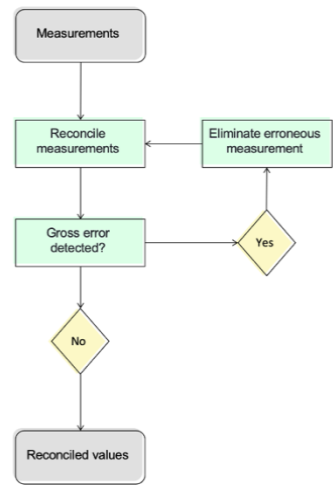

Fig. 2 - Example workflow of gross error detection and elimination

There are multiple ways of performing GED, where all of these have in common that they involve statistical tests and require data redundancy to detect the measurements that are out-of-spec. In this paper, we narrow the scope to include only the *maximum power measurement* *test*. Although known to produce many false positive detections, it is quite favorable in terms of its simplicity, computational efficiency, and ability to locate the gross errors [8]. Even though we have presented DVR in relation to any process industry, the sections that follow delve into the field of upstream petroleum production, where short-term optimization and, consequently, computational efficiency can be of the essence. For a more comprehensive overview of various outlier detection algorithms, the reader is referred to [8] and [10].

In the measurement test, we consider the vector of measurement adjustments, $a = y - C\hat{y}$, that is the difference between the measurements and the reconciled values. Substituting $\hat{y}$ with the solution given by Equation (3) leads to the following equation:

$$a = y - CRy = (I - CR)y \quad (4)$$

Premultiplying $a$ by $\Sigma^{-1}$ results in a vector of test statistics with maximum power for detecting a single gross error:

$$d = \Sigma^{-1}a = \Sigma^{-1}(I - CR)y \quad (5)$$

Under the assumption that there are no systematic errors, these test statistics follow a normal distribution, $d \sim N(0, W)$, where

$$W = Cov(d) = \Sigma^{-1}(I - CR)\Sigma(I - CR)^T\Sigma^{-1} \quad (6)$$

For any measurement $y_i$ we can then calculate the normalized test statistic $z_i$ as demonstrated in Equation (7).

$$z_i = \frac{d_i}{\sqrt{W_{ii}}}, i = 1, \ldots, N \quad (7)$$

The normalized test statistics follow a standard normal distribution, $N(0,1)$, when there is no gross error [8], [13]. This is considered to be our null hypothesis, $H_0$. The measurement test involves $N$ univariate tests. Applying $\alpha$ as the level of significance for the tests, a gross error is detected if any of the test statistics exceed the test criterion given in Equation (8).

$$\alpha_{z/2} = \Phi^{-1}\left(1 - \frac{\alpha}{2}\right) \quad (8)$$

Here, $\Phi$ is the cumulative distribution function of the normal distribution. The test criterion satisfies

$$P\left(-z_{\alpha/2} \leq Z \leq z_{\alpha/2}\right) = 1 - \alpha \quad (9)$$

That is, we reject $H_0$ if, for any $i = 1, \ldots, N$, $|z_i| > z_{\alpha/2}$. The sensitivity of the test can be adjusted using $\alpha$ to reduce the number of false positive detections.

## 3    PETROLEUM PRODUCTION: FLOW RATE MEASUREMENTS

Although the theory behind the proposed framework works for several process industries, it has been tried and tested for the upstream petroleum industry. Taking inspiration from Fig. 1, nodes 1 through $M - 1$ could then be wells, node $M$ may represent a separator, and measurement $y = (y_1, \ldots, y_N)$ could be oil flow rate measurements of varying types. Flow rate measurements can be split into two main categories: measurements from a physical instrument and estimates from a *virtual flow meter* (VFM). The former category includes measurements from a *multiphase flow meter* (MPFM) and topside fiscal measurements. The latter one involves mechanistic models, which are usually created using multiphase flow simulators, and data-driven models, where the VFM is developed based on the production data using machine learning models.

An MPFM is a device used to measure oil, gas, and water flow rates in a pipeline without separating the phases. The meters are usually installed at the wellhead to monitor the well's real-time flow rates or topside to monitor the commingled line production. The flow rates are calculated indirectly using supplementary measurements of fluid phase properties such as velocities and phase fractions inside the device [14]-[15]. MPFMs apply several types of technologies that enable them to estimate oil, gas, and water flow rates, such as acoustic attenuation, impedance, and gamma densitometers [14]. An asset with MPFMs installed presents an advantage from an operational perspective as it provides real-time information about the well's flow rates. However, the devices' measurement accuracy may decrease significantly outside of their optimal operating range or when subject to flow assurance issues [16].

The second type of physical meter is the topside fiscal meter, which is usually responsible for the connection between purchase and sale and the calculation of taxes and royalties [17]. Consequently, the meter system accuracy should meet strict accuracy requirements to ensure fair transactions and compliance with regulatory standards.

The mechanistic VFMs use multiphase flow simulators to model the flow in the wellbore, pipelines, and production chokes. The multiphase flow model consists of a set of models, such as fluid models and thermal and pressure drop models, that are applied together with pressure and temperature values to derive estimates of the flow rates. The production system can be modelled as a whole from the reservoir to the processing facility, or it can be separated into submodels depending on the available measurement data. First principles models are currently used in most commercial virtual flow metering systems, such as Prosper, LedaFlow, Olga, and FlowManager. Similar to MPFMs, mechanistic VFMs strongly rely on fluid properties. Due to uncertain subsurface fluid properties and complex multiphase flow dynamics, it can be challenging to achieve the required precision in real-time using this type of approach [18]-[19]. Lastly, to achieve precise estimates from a mechanistic VFM, a tuning algorithm adjusts the model parameters to minimize the mismatch between the model predictions and real measurements [20].

Data-driven VFMs are often machine learning models trained on production data to estimate flow rates continuously. A typical workflow for data-driven VFMs consists of three main steps: the data preprocessing, the model development, including the training and validation procedures, and lastly, the real-time predictions of new data using the validated model. Although it is not a new concept, the adoption of data-driven VFMs is still quite low in the petroleum industry [6]. This behavior can be motivated by some challenges that arise when working with data-hungry methodologies in this particular industry, such as the limited availability of high-quality historical data and the need for process insights in advanced feature engineering. Furthermore, data-driven VFMs are the most precise when applied to data within or near the training data range. If this is not the case, calibration and re-training on data from the new and unseen domain is required [7]. However, one of the advantages of this type of model is that there is no need to develop a detailed physical model, which could make it hard to find a solution numerically. Additionally, it does not require a deep understanding of the underlying physics behind the system, and it has a lower computational cost when compared to its mechanistic counterparts.

One common denominator between the mentioned measurements is that all of them can be calibrated in one way or another. In fact, for the proposed framework to derive a valid reconciliation, it is often beneficial that some form of calibration takes place from time to time. Calibration is to correct the error between a measured or estimated value and a higher accuracy reference value [17], [21]. Fluid properties and well test data is commonly used when calibrating virtual flow meters [7]. A method for calibrating MPFMs is described by Gustavsen et al. [22].

## 4     THE FRAMEWORK IN PRACTICE: GROSS ERROR DETECTION

### 4.1    The proposed framework

The proposed framework is an industry-tested service for reconciliation and allocation. It is a fully automated system that exploits the information in the typically large amount of production data available in the oil and gas industry. As mentioned above, the solution consists of four modules: data processing, uncertainty estimation, reconciliation, and GED.

The data processing module is essential to retrieve, structure, and format the data in real time. Additionally, by incorporating logic to filter out or correct invalid sensor measurements, it also handles some of the data issues that are present in the petroleum production data, such as poor data quality, from the outset. The second module estimates the uncertainties of the measurements, which are required in the DVR problem. The derived uncertainties are live and condition-based and are based on the measurements' performance against historical well tests, signal processing and a priori information, such as qualified subjective assumptions from industry professionals and data sheets. The DVR problem is implemented and solved through the reconciliation module, before the last module applies GED through the maximum power measurement test for production systems with redundancy in their flow rate measurements. A flow chart of the proposed framework including the four modules is illustrated in Fig 3. Here it is visualized that the gross error detections are primarily sent back to the asset's data base. These can be passively applied to monitor drifting measurement errors, faulty measurements, and equipment failures. However, the figure also demonstrates that they can take on a more active role in terms of gross error elimination, where the reconciled flow rates can be recalculated without the measurements that trigger the gross error detections. This typically necessitates additional logic to ensure that the solution, to the best of its ability, only acts on true positive detections and still has a sufficient amount of data after the measurements are removed for the reconciliation to take place. For instance, a simple algorithm could be to only remove the measurements that both trigger a detection and indicate that the well is producing under a certain threshold when it is expected to produce higher volumes.

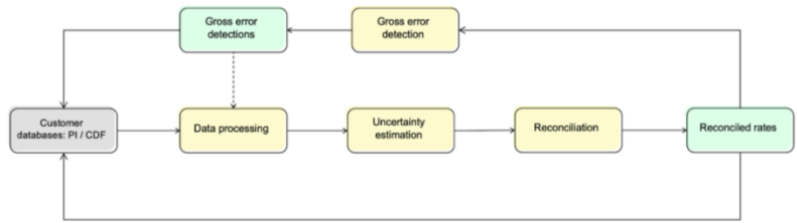

Fig. 3 - Flow chart of the proposed framework

It is ideal for the proposed allocation framework to work on top of data sets with measurement redundancy, and often these involve a combination of the various measurement types described in Section 3. It is, however, important to be aware of how the measurements differ and also handle them accordingly. In the following, we focus on some examples of real-life detections from the GED module and discuss how they came about and their implications for the reconciled flow rates.

### 4.2    Study of some gross error detections

In the following figures, we visualize the reconciled oil flow rates that are derived by the proposed solution with a dark grey and dashed line. In the figure legends, they are abbreviated by "RECON". Also shown are the flow rate measurements that are input to the reconciliation problem with accompanying gross error detections, that can be distinguished by the "x"-markers. We apply "o"-markers otherwise. Furthermore, light red shaded areas emphasize the time intervals within which at least one gross error is detected. For this specific study, the framework is implemented for an oil field on the Norwegian continental shelf. Each well has three different ways of measuring or estimating the oil flow rate - an MPFM, a data-driven VFM, and a mechanistic VFM. In the subsequent figures, these are denoted by "MPFM", "DD-VFM", and "M-VFM", respectively, and represented by the blue, pink, and green lines. The methodology is applied for 24-hour periods, estimating the daily allocation for each well. Hence, the flow rates are shown as average values over each such period, resulting in curves with a "step function" look. In some of the examples we also visualize the measurement data in its raw form, in which case we refer to them as "MPFM raw", "DD-VFM raw", and "M-VFM raw" in the figures.

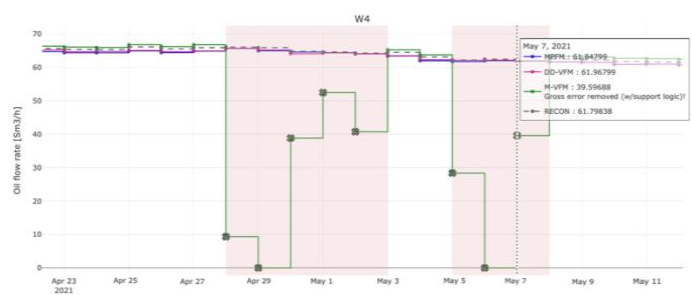

Fig. 4 - Successful gross error detections and eliminations in times of VFM failure for well W4

The first examples of GED are demonstrated in Fig. 4 between April 28$^{th}$ and May 8$^{th}$, 2021. In this period, it is quite obvious that the mechanistic VFM is, on and off, producing measurements of suspicious character for well W4; the average measurements are far below those from the MPFM and data-driven VFM. Looking at the measurements from the mechanistic VFM in its raw form in Fig. 5, it becomes apparent that some sort of failure has occurred. Between the two given dates, there are three smaller periods where the VFM is continuously estimating no flow at all, even though the well is known to be open and producing. This also explains why the VFM's average values in Fig. 4 jumps up and down depending on where each 24-hour period begins and ends.

This is a typical case where we can successfully apply gross error elimination in a very direct manner. When the framework is applied with all measurements included, the reconciled flow rates are wrongfully drawn towards the faulty measurements. As a result, gross errors are detected not only for the malfunctioning VFM, but for the other sensors as well. In other words, the systematic error is located to the most probable well, but not pinpointed to the sensor that is to blame. However, with supporting logic in place, it is possible to single out the mechanistic VFM's measurements as the truly erroneous ones and perform the reconciliation without them as long as they exhibit out-of-spec behavior. In Fig. 4, we observe that the reconciled flow rates in grey remain unaffected by the faulty measurements given by the mechanistic VFM between April 28[th] and May 8[th], demonstrating how the proposed framework is able to both detect and remove such a gross error to mitigate the potential misallocation.

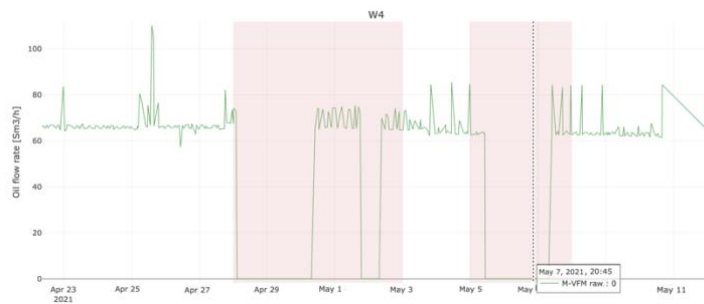

Fig. 5 - Live estimates from the mechanistic VFM during periods of failure for well W4

Another similar example, yet very different in the way the gross errors play out, is the one visualized in Fig. 6. In this figure, between November 2[nd] and November 16[th], 2022, we observe that the reconciled flow rates for well W8 end up somewhere in between the input measurements and gross error detections are occasionally triggered for all of these at the same time - a situation that was briefly described in the last paragraph and categorized as a probable localization of the systematic error. Whereas the task of identifying the erroneous sensor was quite simple in the last example, this time around it is not. In hindsight, we know that the mechanistic VFM was subject to imprecise modelling at the time; The modelled region did not include the operation points that were seen during this period and the VFM was later calibrated to make up for this. However, in real-time, it is not straightforward to know which of the sensors are performing out-of-spec based on the production data alone, especially when gross error detections are triggered for all input measurements. In this particular case, the supporting logic is not able to distinguish the faulty sensor, but delving into the normalized test statistics and ranking them gives us more useful information. On November 13$^{th}$, the test statistics were 8.7, 4.1 and 5.1 for the mechanistic VFM, MPFM and data-driven VFM, respectively. With the highest test statistic, the mechanistic VFM can be correctly suggested to be the erroneous sensor. In the current implementation, this information is not applied to automatically eliminate a gross error from the reconciliation problem and all measurements are therefore still part of it on November 13$^{th}$, though with the potential to be manually removed by a user that could be informed of such details. An alternative implementation, however, could use the test statistics more proactively. This example highlights some of the limitations related to the maximum power measurement test being applied for GED in its simplest form. Nevertheless, although the method is not able to pinpoint the erroneous sensor by the triggered detections alone, it still makes us aware that something is not entirely right at a specific location and can further point us in the right direction if we delve into the details. At the least, these are valuable insights for when production engineers go looking for sensors that might need a service or recalibrating.

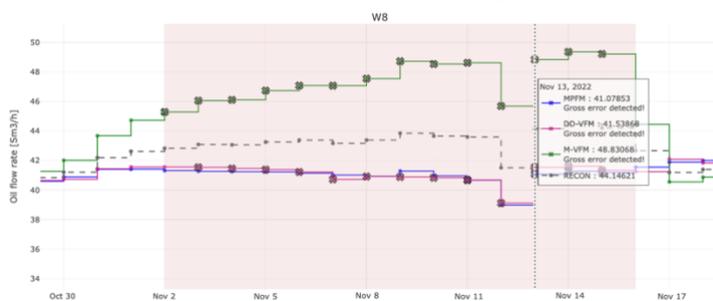

Fig. 6 - Successful localization of gross errors in times of inaccurate modelling for well W8

Lastly, we illustrate a case of inaccurate modelling by the data-driven VFM in Fig. 7. Quite similar to the example shown in Fig. 6, during January 2$^{nd}$, 2023, the production data for well W4 enters a domain that is infrequently seen in the data set the data-driven VFM is trained on. Hence, it has a limited data foundation to base its predictions on and, consequently, fails to accurately predict the oil flow rate. Although a gross error is correctly detected during the relevant time period, it comes in a bit late the following day. The sensitivity of the algorithm for GED is based on a trade-off between capturing all possible systematic errors and triggering only true positive detections. The fact that we are not able to detect the systematic error for the days before and after January 3$^d$ is a result of this trade-off, applying rather strict significance levels in our implementation for the field. We use values that are 0.01 to 0.0001 depending on the measurement type. Furthermore, the gross error is merely detected, and not eliminated, following a conservative philosophy with regards to automatically removing measurements, that we have agreed on with the production engineers of the asset. However, in an alternative implementation, this detection would be fairly straightforward to actually act on since it is triggered for only one of the measurements.

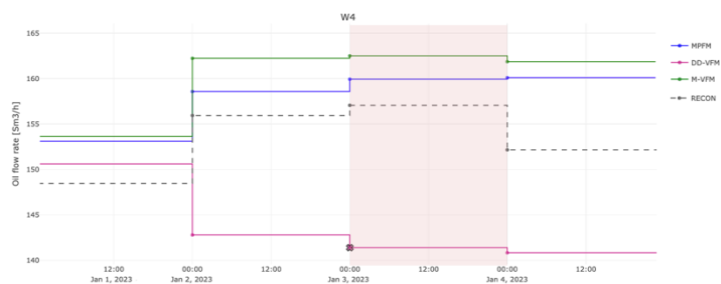

Fig. 7 - Successful gross error detection in times of inaccurate modelling for well W4

It should be noted that this study only looks into a subset of examples to demonstrate both the usability and limitations of our solution for GED within our reconciliation framework. Nevertheless, we believe that both a passive and active application of the gross error detections can, in the end, greatly improve the validity of the allocation in general terms. Yet, proving this statistically remains a research question

## 5  FINAL REMARKS

Within the oil and gas sector, allocation is the task of accurately distributing the total hydrocarbon production to the contributing streams to ensure a justifiable revenue split between stakeholders. Furthermore, failure to appropriately allocate well flow rates negatively affects reservoir simulation, drawdown strategy, and future infill campaigns.

In this paper, a framework for reconciliation and allocation is presented as an alternative solution to conventional allocation methods commonly found in the industry. It is an industry-tested solution that exploits the high-quality information that lies in the production data, such as quantifiable uncertainties, well tests, measurement redundancy, and detectable MPFM or VFM errors. The proposed framework implements DVR for reconciliation, where uncertainties of the measurements are used to mitigate the imbalance in the measured flow rates.

Given that there is measurement redundancy, we apply a module for GED to limit the systematic errors in the set of measurements, which deteriorates the solution from DVR if present. Although the study from Section 4 mainly demonstrates how our solution for GED is able to locate sensor failure and inaccurate modelling, it is a versatile method that can capture any type of systematic error. The gross error detections from the module can be acted on either actively or passively. With the help of additional logic, the detected gross errors can be automatically eliminated from the reconciliation to ensure a more valid outcome. They also form valuable insight for general monitoring and decision making processes, such as whether one should perform a well test and subsequently calibrate a seemingly erroneous meter. Either way, the GED module makes for a more intelligent reconciliation framework that has the potential to achieve a truly valid production allocation.

## 6  NOTATION

| | | | |
|---|---|---|---|
| $A$ | constraint matrix | $\sigma^2$ | variance |
| $a$ | vector of measurement adjustment | $\Sigma$ | variance matrix for the measurements |
| $\alpha$ | level of significance | $W$ | covariance matrix for the test statistics |
| $C$ | measurement matrix | | |
| $d$ | vector of test statistics | $W_{ii}$ | variance for the test statistic for measurement $i$ |
| $d_i$ | test statistic for measurement $i$ | | |
| $\epsilon$ | random error | $y^*$ | vector of true values |
| $H_0$ | null hypothesis | $y$ | vector of measured values or measurements |
| $M$ | number of nodes | | |
| $N$ | number of measurements | $y_i$ | measurement $i$ |
| $\Phi$ | cumulative distribution function of a normal distribution | $\hat{y}$ | vector of allocated values |
| | | $z_i$ | normalized test statistic for measurement $i$ |
| $Q$ | number of constraints | $z_{\alpha/2}$ | test criterion |

## 7  REFERENCES

[1]  A. Amin. *Using Measurement Uncertainty in Production Allocation*, Upstream Production Measurement Forum (2016).

[2]    S. Kanshio. *A review of hydrocarbon allocation methods in the upstream oil and gas industry*, Journal of Petroleum Science and Engineering, 184 (2020).

[3]    A. Wee. *Using multiphase and wet gas meters for revenue allocation*, North Sea Flow Measurement Workshop (2014).

[4]    A. Pobitzer, A. M. Skålvik and R. N. Bjørk. *Allocation system setup optimization in a cost-benefit perspective*, J. Petrol. Sci. Eng. 147, 707–717 (2016). DOI: http://dx.doi.org/10.1016/j.petrol.2016.08.025.

[5]    D. S. van Putten, L. van Luijk, I. S. Wahab and S. F. Johari. *Assessment of allocation systems: combining data validation & reconciliation scheme and PVT simulations*, North Sea Flow Measurement Workshop (2019). URL: https://www.tekna.no/contentassets/eac487e9aabb4773a1337fda3c28b70e/06-assesment-of-alloction-systems_dnvgl_dennis-putten.pdf.

[6]    K. Balaji, M. Rabiei, V. Suicmez, C. H. Canbaz, Z. Agharzeyva, S. Tek and U. Bulut. *Status of Data-Driven Methods and their Applications in Oil and Gas Industry Introduction to Data Driven Methods*, Society of Petroleum Engineers (2018).

[7]    T. Bikmukhametov and J. Jäschke. *First principles and machine learning virtual flow metering: A literature review*, Journal of Petroleum Science and Engineering, vol. 184 (2020). DOI: http://dx.doi.org/10.1016/j.petrol.2019.106487.

[8]    S. Narasimhan and C. Jordache. *Data Reconciliation & Gross Error Detection: An Intelligent Use of Process Data*, Gulf Publishing Company, Houston (2000).

[9]    M. M. Câmara, R. M. Soares, T. Feital, T. K. Anzai, F. C. Diehl, P. H. Thomson and J. C. Pinto. *Numerical Aspects of Data Reconciliation in Industrial Applications*, Processes vol. 5 no. 4 (2017). DOI: http://dx.doi.org/10.3390/pr5040056.

[10]   T. S. Badings and D. S. van Putten. *Data validation and reconciliation for error correction and gross error detection in multiphase allocation systems*, Journal of Petroleum Science and Engineering, 195 (2020).

[11]   C. Benqlilou. *Data Reconciliation as a Framework for Chemical Processes Optimization and Control*, ProQuest dissertations and theses (2004). URL: https://upcommons.upc.edu/handle/2117/93748.

[12]   M. J. Bagajewicz and Q. Jiang. *Gross error modeling and detection in plant linear dynamic reconciliation*, Comput. Chem. Eng. 22, 1789–1809 (1998). DOI: http://dx.doi.org/10.1016/S0098-1354(98)00248-8.

[13]   A. C. Tamhane and R. S. Mah. *Data reconciliation and gross error detection in chemical process networks*, Technometrics 27, 409–422 (1985). DOI: http://dx.doi.org/10.1080/00401706.1985.10488080.

[14] G. Falcone, G. Hewitt, C. Alimonti and B. Harrison. *Multiphase flow metering: current trends and future developments*, SPE Annual Technical Conference and Exhibition (2001). DOI: https://doi.org/10.2118/71474-MS.

[15] A. Gryzlov. *Model-based estimation of multi-phase flows in horizontal wells*, Delft 5–15 PhD dissertation (2011). ISBN: 978-90-6464-455-9.

[16] C. Marshall and A. Thomas. *Maximising economic recovery - a review of well test procedures in the North Sea*, SPE Offshore Europe Conference and Exhibition (2015).

[17] Norwegian Petroleum Directorate (2001). *Regulations relating to measurement of petroleum for fiscal purposes and for calculation of CO2-tax (the measurement regulations)*. URL: https://www.npd.no/en/regulations/regulations/measurement-of-petroleum-for-fiscal-purposes-and-for-calculation-of-co2-tax/. (Accessed 25.09.2023).

[18] J. D. Jansen. *Nodal Analysis of Oil and Gas wells. Theory and Numerical Implementation*, TU Delft (2015).

[19] J. P. Couput, N. Laiani and V. Richon. *Operational Experience with Virtual Flow Measurement Technology*, 35th International North Sea Flow Measurement Workshop (2017).

[20] K. Holmås and A. Løvli. *FlowManager Dynamic: a multiphase flow simulator for online surveillance, optimization and prediction of subsea oil and gas production*, 15th International Conference on Multiphase Production (2011).

[21] Y. Choi and S. Yoon. *Virtual sensor-assisted in situ sensor calibration in operational HVAC systems*, Building and Environment, vol. 181 (2020). DOI: https://doi.org/10.1016/j.buildenv.2020.107079.

[22] S. H. H. Gustavsen, T. Selanger and T. Renstrøm. *Parallel calibration of multiphase flow meters vs separator*, North Sea Flow Measurement Workshop (2019).